\documentclass[twocolumn,trackchanges]{aastex701}

\usepackage{amsmath,amstext,amssymb}
\let\tablenum\relax
\usepackage{siunitx}
\usepackage[T1]{fontenc}
\usepackage{times}
\usepackage{apjfonts} 
\usepackage[figure,figure*]{hypcap}
\usepackage{natbib}
\usepackage{enumerate}
\usepackage[flushleft]{threeparttable}
\usepackage{booktabs}
\usepackage{graphicx}
\usepackage{multirow}
\usepackage[encapsulated]{CJK}
\usepackage[caption=false]{subfig}
\usepackage{hyperref}
\usepackage{needspace}

\usepackage{soul} 
\usepackage{xspace}
\usepackage{xifthen}
\usepackage{eso-pic}

\usepackage{xcolor}
\usepackage{braket}

\newcommand{\spacer}{\needspace{1\baselineskip}}

\newcommand{\numpy}{\code{NumPy}}
\newcommand{\scipy}{\code{SciPy}}
\newcommand{\matplotlib}{\code{matplotlib}}

\newcommand{\cigale}{\code{CIGALE}}

\definecolor{forestgreen}{HTML}{228B22}
\definecolor{urlblue}{HTML}{000000}

\mathchardef\mhyphen="2D

\newlength{\dhatheight}

\newcommand{\code}[1]{\texttt{#1}\xspace}

\newcommand{\dex}{\unit{dex}}
\newcommand{\yr}{\unit{yr}}

\newcommand{\Myr}{\unit{Myr}}

\newcommand{\cm}{\unit{cm}}
\newcommand{\km}{\unit{km}}

\newcommand{\second}{\unit{s}}

\newcommand{\erg}{\unit{erg}}
\newcommand{\ergs}{\erg\;\second^{-1}}

\newcommand{\Msun}{M_\odot}
\newcommand{\Msunyr}{\Msun\yr^{-1}}

\newcommand{\Lsun}{L_\odot}

\newcommand{\e}{\unit{e^{-}}}

\newcommand{\tabref}[1]{Table~\ref{tab:#1}}

\newcommand{\figref}[1]{Figure~\ref{fig:#1}}

\newcommand{\bandvar}[2][]{%
  \ifthenelse{\isempty{#1}}{\var{#2}}{\var{#2\_#1}}%
}

\newcommand{\var}[1]{\ensuremath{\texttt{\MakeUppercase{#1}}}\xspace}

\providecommand\physrep{\ref@jnl{Phys.~Rep.}}%
\providecommand\apjs{\ref@jnl{ApJS}}%
\providecommand{\jcap}{\ref@jnl{JCAP}}%

\newcommand{\ctagn}{{CT AGN}\xspace}
\newcommand{\one}{{\textit{el Ascua}}\xspace}

\received{XXX}
\revised{XXX}
\accepted{XXX}
\submitjournal{\apjl}

\shorttitle{A distant and extremely powerful \ctagn}
\shortauthors{A. P. Cloonan et al.}

\begin{document}

\title{\textit{El Ascua}: An ultra-luminous, radio-loud, Compton-thick AGN at $z\sim 3.8$ hiding in plain sight revealed by MINERVA multi-band MIRI observations}

\author[0000-0001-9978-2601]{Aidan P. Cloonan}\thanks{NSF Graduate Research Fellow}
\affiliation{Department of Astronomy, University of Massachusetts, Amherst, MA 01003, USA}
\email[show]{apcloonan@umass.edu}

\author[orcid=0009-0002-2209-4813,gname=Edgar,sname=Vidal]{Edgar P. Vidal}
\affiliation{Department of Physics and Astronomy, Tufts University, 574 Boston Avenue, Suite 304, Medford, MA 02155, USA}
\email{edgar.vidal@tufts.edu}

\author[0000-0001-8592-2706]{Alexandra Pope}
\affiliation{Department of Astronomy, University of Massachusetts, Amherst, MA 01003, USA}
\email{pope@astro.umass.edu}

\author[0000-0002-1917-1200]{Anna Sajina}
\affiliation{Department of Physics and Astronomy, Tufts University, 574 Boston Avenue, Suite 304, Medford, MA 02155, USA}
\affiliation{David A. Dunlap Department of Astronomy \& Astrophysics, University of Toronto, 50 St. George Street, Toronto, ON M5S 3H4, Canada}
\email{annie.sajina@gmail.com}

\author[0000-0003-3243-9969]{Nicholas S. Martis}
\affiliation{University of Ljubljana, Faculty of Mathematics and Physics, Jadranska ulica 19, SI-1000 Ljubljana, Slovenia}
\email{nicholas.martis@fmf.uni-lj.si}

\author[0000-0003-3983-5438]{Yoshihisa Asada}
\thanks{Dunlap Fellow}
\affiliation{David A. Dunlap Department of Astronomy \& Astrophysics, University of Toronto, 50 St. George Street, Toronto, ON M5S 3H4, Canada}
\affiliation{Waseda Research Institute for Science and Engineering, Faculty of Science and Engineering, Waseda University, 3-4-1 Okubo, Shinjuku, Tokyo 169-8555, Japan}
\email{yoshi.asada@utoronto.ca}

\author[0000-0003-3881-1397]{Olivia R. Cooper}
\thanks{NSF Astronomy and Astrophysics Postdoctoral Fellow}
\affiliation{Department for Astrophysical \& Planetary Science, University of Colorado, Boulder, CO 80309, USA}
\email{olivia.cooper@colorado.edu}

\author[0000-0002-9330-9108]{Adam Muzzin}
\affiliation{Department of Physics and Astronomy, York University, 4700 Keele St., Toronto, Ontario, M3J 1P3, Canada}
\email{muzzin@yorku.ca}

\author[0000-0002-8909-8782]{Stacey Alberts}
\affiliation{AURA for the European Space Agency (ESA), Space Telescope Science Institute, 3700 San Martin Dr., Baltimore, MD 21218, USA}
\email{salberts@stsci.edu}

\author[0000-0002-0243-6575]{Jacqueline Antwi-Danso}
\thanks{Banting Postdoctoral Fellow}
\affiliation{David A. Dunlap Department of Astronomy \& Astrophysics, University of Toronto, 50 St. George Street, Toronto, ON M5S 3H4, Canada}
\affiliation{Department of Astronomy, University of Massachusetts, Amherst, MA 01003, USA}
\email{j.antwidanso@utoronto.ca}

\author[0000-0003-2680-005X]{Gabriel Brammer}
\affiliation{Cosmic Dawn Center (DAWN), Copenhagen, Denmark}
\affiliation{Niels Bohr Institute, University of Copenhagen, Jagtvej 128, DK-2200 Copenhagen N, Denmark}
\email{gabriel.brammer@nbi.ku.dk}

\author[0000-0002-7031-2865]{Sam E. Cutler}
\affiliation{Department of Physics and Astronomy, Tufts University, 574 Boston Avenue, Suite 304, Medford, MA 02155, USA}
\email{sam.cutler@tufts.edu}

\author[0000-0001-9298-3523]{Kartheik G. Iyer}
\thanks{NHFP Hubble Fellow}
\affiliation{Columbia Astrophysics Laboratory, Columbia University, 550 West 120th Street, New York, NY 10027, USA}
\email{kartheikiyer@gmail.com}

\author[0000-0002-2057-5376]{Ivo Labb\'e}
\affiliation{Centre for Astrophysics and Supercomputing, Swinburne University of Technology, Melbourne, VIC 3122, Australia}
\email{ivolabbe@gmail.com}

\author[0000-0001-9002-3502]{Danilo Marchesini}
\affiliation{Department of Physics and Astronomy, Tufts University, 574 Boston Avenue, Suite 304, Medford, MA 02155, USA}
\email{danilo.marchesini@tufts.edu}

\author[0000-0001-8367-6265]{Tim B. Miller}
\affiliation{Center for Interdisciplinary Exploration and Research in Astrophysics (CIERA), Northwestern University, 1800 Sherman Ave, Evanston, IL 60201, USA}
\email{timothy.miller@northwestern.edu}

\author[0000-0001-7300-9450]{Ikki Mitsuhashi}
\affiliation{Department for Astrophysical \& Planetary Science, University of Colorado, Boulder, CO 80309, USA}
\email{ikki0913astr@gmail.com}

\author[0000-0002-6265-2675]{Luke Robbins}
\affiliation{Department of Physics and Astronomy, Tufts University, 574 Boston Avenue, Suite 304, Medford, MA 02155, USA}
\email{andrew.robbins@tufts.edu}

\author[0000-0001-8830-2166]{Ghassan T. E. Sarrouh}
\affiliation{Department of Physics and Astronomy, York University, 4700 Keele St., Toronto, Ontario, M3J 1P3, Canada}
\email{gsarrouh@yorku.ca}

\author[0000-0003-1078-9706]{Monu Sharma}
\affiliation{Departament d'Astronomia i Astrofisica, Universitat de Valencia, C. Dr. Moliner 50, E-46100 Burjassot, Valencia, Spain}
\email{mnushv@gmail.com}

\author[0000-0001-7768-5309]{Mauro Stefanon}
\affiliation{Departament d'Astronomia i Astrofisica, Universitat de Valencia, C. Dr. Moliner 50, E-46100 Burjassot, Valencia, Spain}
\affiliation{Unidad Asociada CSIC ``Grupo de Astrofisica Extragalactica y Cosmologi'' (Instituto de Fisica de Cantabria - Universitat de Valencia)}
\email{mauro.stefanon@uv.es}

\author[0000-0002-1714-1905]{Katherine A. Suess}
\affiliation{Department for Astrophysical \& Planetary Science, University of Colorado, Boulder, CO 80309, USA}
\email{kasu8993@colorado.edu}

\author[0000-0001-7160-3632]{Katherine E. Whitaker}
\affiliation{Department of Astronomy, University of Massachusetts, Amherst, MA 01003, USA}
\affiliation{Cosmic Dawn Center (DAWN), Copenhagen, Denmark}
\email{kwhitaker@astro.umass.edu}

\author[0000-0002-4201-7367]{Chris J. Willott}
\affiliation{National Research Council of Canada, Herzberg Astronomy \& Astrophysics Research Centre, 5071 West Saanich Road, Victoria, BC, V9E 2E7, Canada}
\email{chris.willott@nrc.ca}

\author[0000-0001-8460-1564]{Pratika Dayal}
\affiliation{Canadian Institute for Theoretical Astrophysics, 60 St George St, University of Toronto, Toronto, ON M5S 3H8, Canada}
\affiliation{David A. Dunlap Department of Astronomy \& Astrophysics, University of Toronto, 50 St. George Street, Toronto, ON M5S 3H4, Canada}
\affiliation{Department of Physics, 60 St. George Street, University of Toronto, Toronto, ON M5S 3H8, Canada}
\email{pdayal@cita.utoronto.ca} 

\author[0000-0002-2380-9801]{Anna de Graaff}
\affiliation{Max-Planck-Institut f\"ur Astronomie, K\"onigstuhl 17, D-69117, Heidelberg, Germany}
\email{degraaff@mpia.de}

\author[0000-0001-6003-0541]{Ben Forrest}
\affiliation{Department for Astrophysical \& Planetary Science, University of Colorado, Boulder, CO 80309, USA}
\email{ben.forrest@colorado.edu}

\author[0000-0002-3254-9044]{Karl Glazebrook}
\affiliation{Centre for Astrophysics and Supercomputing, Swinburne University of Technology, Melbourne, VIC 3122, Australia}
\email{kglazebrook@swin.edu.au}

\author[0000-0002-5612-3427]{Jenny E. Greene}
\affiliation{Department of Astrophysical Sciences, Princeton University, 4 Ivy Lane, Princeton, NJ 08544, USA}
\email{crispygreene@gmail.com}

\author[0009-0006-5313-3810]{Kelsey Hargraves}
\affiliation{School of Chemistry and Physics, 2 George St, Queensland University of Technology, Brisbane, QLD 4000, Australia}
\email{kelsey.hargraves@connect.qut.edu.au}

\author[0000-0001-5916-5641]{William W.H. Jarvis}
\affiliation{Department of Astronomy, University of Massachusetts, Amherst, MA 01003, USA}
\email{wjarvis@umass.edu}

\author[orcid=0000-0002-7847-1107]{Daniel J. Lyon}
\affiliation{Centre for Astrophysics and Supercomputing, Swinburne University of Technology, Melbourne, VIC 3122, Australia}
\email{djlyon@swin.edu.au}

\author[0000-0002-2446-8770]{Ian McConachie}
\affiliation{Department of Astronomy, University of Wisconsin-Madison, 475 N. Charter St., Madison, WI 53706 USA}
\email{ian.mcconachie@wisc.edu}

\author{Gaël Noirot}
\affiliation{Space Telescope Science Institute, 3700 San Martin Drive, Baltimore, MD 21218, USA}
\email{gnoirot@stsci.edu}

\author[0009-0001-8728-6894]{Veronica Pratt}
\affiliation{Department of Physics and Astronomy, Tufts University, 574 Boston Avenue, Suite 304, Medford, MA 02155, USA}
\email{veronica.pratt@tufts.edu}

\author[0000-0001-9269-5046]{Bingjie Wang}
\thanks{NHFP Hubble Fellow}
\affiliation{Department of Astrophysical Sciences, Princeton University, 4 Ivy Lane, Princeton, NJ 08544, USA}
\email{bjwang@princeton.edu}

\begin{abstract}

Heavily dust-obscured active galactic nuclei (AGN) are crucial for understanding supermassive black hole growth. Using multi-band MIRI photometry from the MINERVA survey, we report the discovery in the UDS field of an extremely mid-infrared bright Compton-thick AGN (CT AGN) at $z \sim 3.8$, \textit{el Ascua} (the Ember). The AGN and/or its host galaxy is detected at 870\,$\mu$m with ALMA and at 1.4\,GHz with the VLA. With UV-to-radio spectral energy distribution modeling, we estimate that the dusty host galaxy has a stellar mass of $\log\left( M_\star / M_\odot \right)=10.82 \pm 0.57$ and a star formation rate of ${\rm{SFR}} =157 \pm 8 \,\Msunyr$, and we confirm the presence of a powerful dusty AGN with an intrinsic accretion luminosity of $\log\left( L_{\rm{acc}} /\ergs \right)=47.12 \pm 0.58$. CT obscuration is revealed by comparing the upper limit on the X-ray luminosity over $2-10 \,{\rm{keV}}$ with the $6 \,{\rm\mu{m}}$ mid-IR AGN luminosity. We additionally find evidence for a radio excess by comparing its total IR-to-radio ratio to those of typical star-forming galaxies and radio-quiet AGN, implying that \textit{el Ascua} could be a rare radio-loud CT AGN. With an extremely high ratio between black hole accretion rate and SFR, the AGN is likely nearing the peak of a deeply obscured phase of black hole growth before AGN feedback blows away the obscuring material. Our work demonstrates the power of MIRI to reveal heavily obscured AGN missing from our supermassive black hole census.

\end{abstract}

\spacer
\section{Introduction}
\label{sec:intro}

Most active galactic nuclei (AGN) are obscured by interstellar gas and dust \citep[e.g.,][]{Hickox18}, and at $z>0$ these obscured AGN are the dominant mode for the growth of supermassive black holes \citep[SMBHs; e.g.,][]{Yang23, Bulichi26}. In particular, Compton-thick AGN (\ctagn), with gas column densities of $N_H \gtrsim 1.5 \times 10^{24}\,\rm cm^{-2}$ attenuating soft X-ray emission (i.e., of energy $<10\; \rm keV$) from hot coronal gas near the accretion disk, are required to match the cosmic X-ray background \citep[e.g.,][]{Gilli07, Draper10}. \ctagn make up a significant fraction of AGN at $z=0$, though the exact number is uncertain with estimates ranging from $\sim 20\%$ to $\sim 50\%$ of AGN \citep{Burlon11, Brightman12, Lanzuisi18, Ananna19}. In the local universe, \ctagn can be found in the heavily buried nuclei of luminous and ultra-luminous infrared galaxies (LIRGs/ULIRGs), with $\sim 20-40\%$ of nearby ULIRGs hosting \ctagn \citep[e.g.,][]{Nardini11, Bauer15, Yamada20, Falstad21, Johnstone25}. At $z \gtrsim 1$, \ctagn account for anywhere from $\sim 33\%$ to $>50\%$ of AGN \citep[e.g.,][]{Bauer10, Brightman12, Buchner15, DelMoro16, Lanzuisi18, Lacy20, Carroll23}, meaning that a similar or even higher fraction of AGN are CT in the distant universe compared to $z=0$ \citep{Treister09}. \ctagn consequently reflect a key evolutionary phase in SMBH growth, and studying their accretion properties, interstellar environments, and host galaxies is critical for comprehensive models of black hole growth, obscuration, fueling, and feedback. 

\ctagn are notoriously difficult to identify and confirm beyond the local universe, as the observed X-ray luminosity at $< 10\; \rm keV$ is tens or hundreds of times fainter than the intrinsic, unabsorbed value \citep{Comastri04}. Furthermore, X-ray surveys are not sufficiently deep at harder rest-frame X-ray energies of $> 10\; \rm keV$ where the obscuring material is more optically thin \citep{Gilli07}. As a result, a significant number of distant \ctagn are missed by both soft and hard X-ray surveys. 

Infrared (IR) observations are therefore crucial for identifying and characterizing \ctagn \citep{Lyu22, Sajina22}, especially at $z>0$ where X-ray constraints are normally unavailable \citep{Gilli14, Vito20}. In lieu of X-ray detections, studies of \ctagn have often paired the weak or non-detected X-ray flux with spectral features in the rest-frame mid-IR (MIR) at $3-30\,\mu{\rm m}$ \citep[e.g,][]{Daddi07, Alexander08, Fiore09, Gandhi09, Goulding12, Gilli14, Lansbury15, Lyu24, Guo25}, where the MIR flux density from an AGN originates from a warm dusty torus enveloping the SMBH and the accretion disk \citep[e.g.,][]{Hickox18, Lyu22}. The MIR features often associated with AGN include the continuum from the warm dusty torus which brightens towards redder MIR wavelengths, forbidden emission lines from ions requiring high ionization potential, and broad absorption features at $9.7\,\mu{\rm m}$ and $18\,\mu{\rm m}$ from silicate dust grains in the foreground of the warm dust continuum \citep{Sajina22}. Spectral energy distribution (SED) modeling has allowed for disentangling the AGN warm dust from polycyclic aromatic hydrocarbons \citep[PAHs;][]{Pope08, Kirkpatrick12, Kirkpatrick15}. PAHs trace star formation and the interstellar medium, and they exhibit unique spectral features in the mid-IR, whereas the warm dusty torus of an AGN shows a strong, red, power-law continuum which can outshine the PAHs. The shape of the mid-IR SED therefore is sensitive to the relative contributions of PAHs and dusty torus emission. However, at $z \gtrsim 1$ studies with \textit{Spitzer} and \textit{Herschel} are generally limited to the brightest obscured AGN \citep[e.g.,][]{Lacy07, Alexander08, Bauer10, Gruppioni13, Carroll23}, rendering larger statistical samples of distant obscured AGN difficult to compile.

The first several years of \textit{JWST} results, especially with the Mid-IR Instrument (MIRI), have dramatically changed the broader landscape of obscured AGN science. The large increase in sensitivity that MIRI provides is pushing population studies of dusty galaxies and AGN to cosmic noon ($z\sim 2$) and beyond \citep[e.g.,][]{Kirkpatrick23, Alberts24, Backhaus26}. With up to eight broadband filters spanning $5-25\,\rm \mu m$, the photometric sampling of the SED allows for more robust decomposition of AGN and PAH emission \citep{Yang23, Lyu24, Bulichi26}, giving stronger constraints on AGN activity across the whole luminosity function. \citet{Lyu24} demonstrate these improvements for \ctagn as part of a broader overview of the power MIRI provides for finding more complete obscured AGN samples. Even without MIRI, \citet{Maiolino25} measure AGN bolometric luminosities from \textit{JWST} near-IR spectra and combine with X-ray brightness to find heavily obscured AGN at $z>4$, where MIRI no longer probes the rest-frame mid-IR. Between \textit{JWST}'s IR capabilities and archival X-ray data, we can now reliably find \ctagn beyond the local universe without relying on rare X-ray detections.

In this \textit{Letter}, we report the discovery of an extremely luminous \ctagn at $z_{\rm phot} \approx 3.77$ in the Ultra-Deep Survey (UDS) extragalactic field \citep{Lawrence07} with the ``Medium-band Imaging with NIRCam to Explore ReVolutionary Astrophysics" (MINERVA) program \citep[GO-7814, PIs: Muzzin, Suess, Marchesini;][]{Muzzin26}. The \ctagn, located at J021725$-$051427, is associated with a sub-millimeter galaxy (SMG) with a strong Band 7 detection in archival ALMA imaging. While it has been flagged previously as a radio source \citep{Simpson06, Zhang25}, its obscured AGN activity\textemdash much less its CT nature\textemdash has not been previously explored. We briefly describe how the source was identified, and we combine the \textit{JWST} data with archival X-ray, optical, IR, and radio data to study its star formation and obscured BH accretion. We name it \one\footnote{``The Ember" in Spanish.} given its red color, strong mid-IR emission from hot dust, and powerful accretion luminosity. In a forthcoming paper (A. P. Cloonan et al. in prep.), we will leverage the unprecedentedly wide area from MINERVA of deep MIRI imaging in $\geq 4$ bands to study a large sample of X-ray-weak and/or CT AGN from $0 < z < 4$. When deriving physical properties, we assume the \textit{Planck} cosmology from \citet{Planck20} with $H_0 = 67.66\;\rm \km\,s^{-1}\,Mpc^{-1}$ and $\Omega_{m,0} = 0.30966$.

\section{Data and Analysis} 
\label{sec:data}
\subsection{MINERVA NIRCam and MIRI}

MINERVA covers several legacy fields with deep imaging in both NIRCam medium bands and in multiple mid-IR filters with MIRI. For the purposes of measuring distant dusty galaxies, this dataset combines fine spectral sampling of the dust-attenuated stellar bump in the rest-optical with multi-band sampling of the rest-frame MIR. We can therefore disentangle PAHs from the dusty torus of the AGN in the mid-IR with significantly greater accuracy and precision on the photometric redshift.

Imaging of the UDS field with the broad-band filters from NIRCam, as well as two MIRI filters (F770W, F1800W) out of four, comes from the Public Release IMaging for Extragalactic Research (PRIMER) treasury program \citep[GO-1837, PI: Dunlop;][]{Donnan24}. MINERVA builds upon PRIMER in UDS by adding eight NIRCam medium bands and two additional MIRI filters (F1280W, F1500W), with a total area of 125 arcmin$^2$ imaged with all four MIRI filters. The archival \textit{HST} data and new NIRCam data are reduced using \texttt{grizli} and following a similar strategy as in \citet{Sarrouh26}. More details of the NIRCam reductions can be found in \citet{Cutler26} and in a subsequent paper (L. Robbins, et al. in prep.). The MIRI data from both PRIMER and MINERVA are reduced with a strategy based on the ``super background" subtraction method introduced by \citet{Alberts24}. The full details of our data reduction strategy will be detailed in a subsequent paper (N. Martis et al. in prep.).

Photometric catalogs are first produced with NIRCam using a modified version of \code{aperpy}, and a brief overview of the method can be found in Section 2.2 of \citet{Cutler26}. Photometric redshifts are fit to the NIRCam photometry with EA\textit{z}Y \citep{Brammer08} with multiple sets of templates; in this paper, we use the redshift catalog created with the \code{sfhz\_blue\_agn} templates. A morphology catalog is constructed from the NIRCam catalog following the method from \citet{Zhang26}, who fit single S\'ersic profiles using the \code{pysersic} code. Finally, photometry in each MIRI filter is measured for each NIRCam source via a fixed-aperture measurement of the color between the MIRI image and the F444W image \citep[PSF-matched to the MIRI resolution; e.g.,][]{Quadri07}. The strategy for full NIRCam$+$MIRI photometric catalogs will be detailed by Y. Asada et al. (in prep.).

\subsection{Ancillary Data}
\label{sec:ancillary}

In our analysis, we leverage archival coverage of the UDS field from \textit{Chandra}, \textit{Spitzer}, the Atacama Large Millimeter Array (ALMA), and the Karl Jansky Very Large Array (VLA). The \textit{HST} and \textit{Spitzer} data provide additional constraints on the UV-optical stellar bump and the mid-IR continuum from AGN and PAHs, while the ALMA coverage in Band 7 (from program 2015.1.01074.S; PI: Inami) places a constraint on the cold dust content and the total IR luminosity. With the X-ray data from the XUDS survey \citep{Kocevski18}, we can verify CT obscuration, and the 1.4\,GHz radio data from VLA \citep{Simpson06, Zhang25} probes the synchrotron emission which can be from star formation or from AGN-driven jets.

We obtain archival \textit{Spitzer} and VLA fluxes from the NASA/IPAC Extragalactic Database. For the ALMA flux at 870\,$\mu$m, we use the \texttt{PyBSDF} code to fit a single 2D Gaussian (with axis ratios of $a/b < 1$ allowed) to the light profile. This surface brightness fit gives a measurement of the total flux observed in an interferometric map, which we include in order to model the panchromatic spectral energy distributions (SEDs).

\subsection{SED Fitting with \cigale}

We use \cigale \citep{Boquien19, Yang20} to model the UV-to-IR SEDs of galaxies in the MINERVA UDS catalog, including the rest-MIR dust emission both from PAHs and the warm dusty torus of an AGN. A full description of the SED modeling and AGN selection in MINERVA with \cigale will be presented by E. Vidal et al. (in prep.), and we briefly outline the approach here. The redshifts are fixed to the best-fit photometric redshifts from the EA\textit{z}Y catalog using the \code{sfhz\_blue\_agn} templates. We adopt the set of priors used by \citet{Yang23, Ling24, Ling26, Hsieh25, Bulichi26, Hamblin26, Lyon26} to jointly fit galaxy$+$AGN emission observed with MIRI. The core ingredients are as follows: a delayed-tau star formation history model, the initial mass function from \citet{Chabrier03}, the nebular emission model from \citet{VillaVelez21}, the dust attenuation model from \citet{Calzetti00}, the ISM dust emission model from \citet{Draine14}, and the \code{SKIRTOR} model for AGN emission from a clumpy torus \citep{Stalevski12, Stalevski16}. These \cigale fits provide mid-IR AGN fractions, AGN luminosities, galaxy stellar masses, and star formation rates. For this paper (see Section 3), we add the photometry from \textit{Spitzer}, ALMA, and the VLA, and then we repeat \cigale fitting for \one to improve the measurement of the IR luminosity and AGN accretion power.

\section{Searching for \ctagn}
\label{sec:selection}

We select \ctagn in the MINERVA UDS catalog by comparing the intrinsic $6\,{\rm \mu m}$ AGN luminosity to the observed $2-10\,{\rm keV}$ X-ray luminosity. The mid-IR luminosity is given directly by the AGN model component in \cigale, corrected for dust attenuation which is significant in the dustiest systems \citep{Stalevski16, Yang20}. The observed X-ray luminosity is uncorrected for gas absorption and is sensitive to $N_H$. The ratio between the two is therefore an excellent tracer for heavy obscuration, especially where direct X-ray detections are not available as is often true for \ctagn at $z>0$ \citep[e.g.,][]{Lansbury15, Hickox18, Lyu24, Uematsu25}. Below, we describe our method for computing upper limits on the $2-10\,{\rm keV}$ X-ray luminosity, and we then outline our selection method for \ctagn. 

\subsection{Upper limits on X-ray luminosity}

We focus on MIRI sources which are undetected in the X-ray maps. Therefore, upper limits on the uncorrected (i.e., absorbed) X-ray luminosity from rest-frame $2-10$\;keV are required, and we use the \texttt{aplimits}\footnote{\url{https://cxc.cfa.harvard.edu/ciao/ahelp/aplimits.html}} routine from the \texttt{CIAO} and \texttt{CALDB} packages for \textit{Chandra} data analysis. In the broad $0.5-7$\;keV band, we estimate the rate of background photon counts averaged over exposure time, telescope area, and solid angle (i.e., in units of $\rm counts\rm\; cm^{-2}\; s^{-1} \;arcsec^{-2}$) by summing the counts over a large source-free region in the X-ray map and dividing that number by the median exposure map value (which has units of $\rm cm^{2}\; s$) and by the solid angle of the region. The \texttt{aplimits} code uses this background rate to measure the minimum number of counts needed in the source aperture for detection, given a source aperture size and desired rates of false positives (where background noise exceeds the detection limit) and false negatives (where a source with exactly the flux upper limit is not flagged). We require a false positive rate of 0.23\%, which corresponds to a $3\sigma$ Gaussian upper limit \citep{Kashyap10}, and a false negative rate of 50\%. We compute with an aperture of radius 3\farcs and a monochromatic energy of 2.3 keV over the broad band, giving us an estimate on the upper limit on the observed-frame $0.5-7$\;keV flux.

We extrapolate these upper limits on flux to the rest-frame $2-10$\;keV range following the method from \citet{Ricci21}. We assume an X-ray spectrum $N(E) \propto E^{-\Gamma}$, where $\Gamma=1.8$ is the photon index. This value for $\Gamma$ is typical for broad-line and unobscured AGN \citep[e.g.,][]{DelMoro17}, meaning that this model roughly corresponds to the maximally allowed spectrum for an unobscured and undetected source. We scale this mock spectrum to match each source's integrated upper limit over the observed $0.5-7$\;keV, and we then integrate over the rest-frame $2-10$\;keV range, and we then convert to a luminosity using the expected photon energy value ($\approx 4.2$\;keV) from the mock spectrum over that energy range.

\subsection{An Extreme \ctagn}

\begin{figure}
    \centering
    \includegraphics[width=\linewidth]{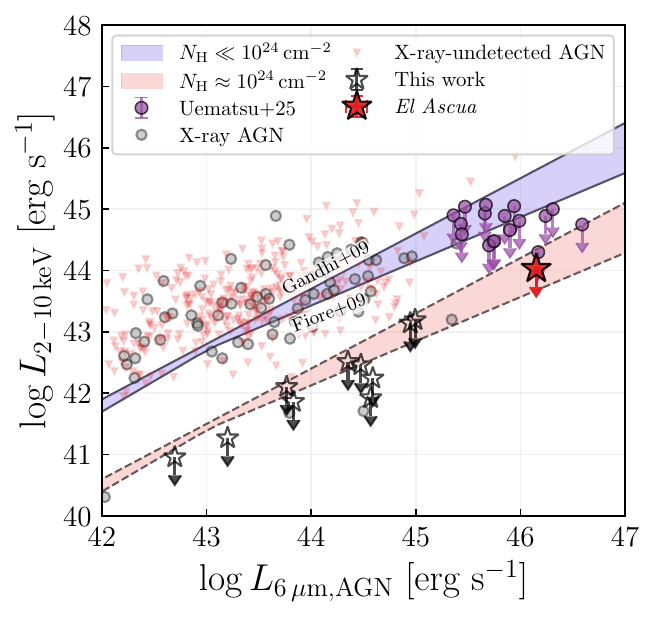}
    \caption{The discovery of a remarkably mid-IR bright ($L_{6\,\rm \mu m,AGN} \gtrsim 10^{46}\,\rm erg\; s^{-1}$) \ctagn candidate (\one; red star) from a larger sample of \ctagn (shown with black stars). This plot shows the observed X-ray luminosity as a function of the intrinsic 6\,$\mu$m AGN luminosity as obtained from SED fitting. The log-linear relation for unobscured (or Compton-thin) AGN is shown in light blue, as a shaded region between two different estimates from \citet{Gandhi09} and \citet{Fiore09}. X-ray detected AGN in UDS which were detected with MINERVA MIRI imaging are shown as gray dots which roughly follows the shaded region for unobscured AGN. 
    The relation for AGN with absorption of $N_H \approx 10^{24}\,\rm cm^{-2}$ is shown in light red, and it corresponds to a normalization factor of $\approx 20$ fainter in the X-ray luminosity \citep{Lansbury15}. A sample of obscured candidate AGN in the COSMOS field are shown as purple dots \citep{Uematsu25}.}
    \label{fig:L6Lx}
\end{figure}

From the initial catalog of 4632 MIRI-detected sources (i.e., $S/N > 5$ in at least 2 MIRI bands) in MINERVA UDS, the criteria which we use to select \ctagn are as follows:
\begin{itemize}
    \item $\chi^2_{\nu} < 3$ for the best-fit SEDs from \cigale, giving 4539 sources.
    \item $f_{\rm AGN} \geq 0.3$ from \cigale, i.e., the AGN contributes at least 30\% of the emission between $3-30\;\rm \mu m$. This selection gives 451 AGN identified via the contribution of the dusty torus to the mid-IR.
    \item The nearest neighbor in the X-ray catalog \citep{Kocevski18} is located more than $3''$ away, giving 375 AGN which are simultaneously bright in the mid-IR and undetected in the X-ray.
    \item The source lies below the \citet{Gandhi09} log-linear relation in $L_{6\, \rm \mu m}-L_{2-10\,\rm keV}$ normalized for CT absorption, giving 11 sources from the MIRI AGN catalog which we identify as \ctagn. The relation from \citet{Fiore09} is slightly more restrictive, and other studies have found that sources between these two relations tend to have CT absorbing material of $N_H \approx 10^{24}\,\rm cm^{-2}$ or slightly denser \citep[e.g.,][]{Lansbury15}. The upper limits on $L_{2-10\,\rm keV}$ are computed as described above.
\end{itemize}
To summarize, we identify \ctagn as MIRI-bright, X-ray-undetected AGN where the upper limit on $L_{6\, \rm \mu m}-L_{2-10\,\rm keV}$ is sufficiently low to place the source below a predicted and empirically verified CT threshold.

From the final CT sample, we find one \ctagn with a mid-IR luminosity of $\log\left( L_{6\, \rm \mu m}/\ergs \right) \gtrsim 46$, much brighter than the others which have $\log\left( L_{6\, \rm \mu m}/\ergs \right) \lesssim 45$. This source is \one at $z_{\rm phot} = 3.78_{-0.03}^{+0.01}$, and we present it and the rest of the CT sample in the $L_{6\, \rm \mu m}-L_{2-10\,\rm keV}$ parameter space in \figref{L6Lx}. Notably, \one has a similar AGN luminosity to a set of AGN-hosting SMGs studied by \citet{Uematsu25} in the COSMOS field. However, the X-ray constraints are stronger for \one than for the COSMOS sample, allowing us to more confidently identify CT obscuration. 

We investigate how the posterior on photometric redshift impacts the upper limit on the X-ray luminosity. If the uncertainty is sufficiently large towards higher redshift, it can bias the limit upward and weaken the constraint on the X-ray obscuration provided by the $L_{6\,\rm \mu m}-L_{2-10\,\rm keV}$ parameter space. For \one, the 95\% confidence interval in redshift is $z_{\rm phot} = 3.78_{-0.18}^{+0.02}$. If we set the redshift to the 97.5\% value of $\approx 3.80$, we find that the upper limit on $L_{2-10\,\rm keV}$ increases by $0.004\,\dex$, remaining below the expected CT threshold shown in \figref{L6Lx}. We therefore conclude that the redshift uncertainty does not affect the interpretation that \one is CT.

For reference, we also plot X-ray AGN in the UDS MINERVA MIRI footprint. We match X-ray sources to the catalog of MIRI sources with best-fit SEDs of $\chi^2_{\nu}  < 3$ by requiring that the X-ray and \textit{JWST} positions are within $1''$ of each other. Qualitatively, we find that the X-ray AGN are consistent with the known Compton-thin scaling relations from \citet{Fiore09} and \citet{Gandhi09}, confirming that our method identifies X-ray-weak sources including \ctagn. Of the X-ray-detected AGN, two are CT given their mid-IR luminosity, leading to a sample size of 13 \ctagn in MINERVA/UDS. In a future paper (A. P. Cloonan et al. in prep.), we will study the full sample of \ctagn and their host galaxy properties in MINERVA.

\section{Source Properties and Discussion}

\begin{deluxetable}{cc|c}
    \tabletypesize{\small}
    \tablewidth{75pt}
    \tablecolumns{3}
    \tablecaption{A number of observables and physical properties of \one, the extreme \ctagn reported in this work. The rows are in order as follows: the right ascension, the declination, the photometric redshift ($z_{\rm phot}$), the AGN fraction over $3-30\,\rm \mu m$ ($f_{\rm AGN}$), the sub-mm flux at 870\,$\mu$m from ALMA, the mid-IR continuum luminosity at $\rm 6\, \mu m$ from the AGN component ($L_{\rm 6\, \mu m, AGN}$) and the total SED ($L_{\rm 6\, \mu m, tot}$), the X-ray luminosity over $\rm 2-10\, keV$ ($L_{\rm 2-10\, keV}$), the accretion power (or the angle-averaged intrinsic disk luminosity; $L_{\rm acc}$), the black hole accretion rate (${\rm BHAR}$) assuming a radiative efficiency of $\epsilon = 0.1$, the host galaxy stellar mass ($M_\star$), the star formation rate (SFR) over the last 100 Myr from the SED fitting, the integrated IR luminosity over $\rm 8-1000\, \mu m$ ($L_{\rm IR}$), the component of $L_{\rm IR}$ from star formation ($L_{\rm IR, SF}$), the radio power at 1.4\,GHz ($P_{\rm 1.4\, GHz}$), and a log-ratio ($q_{\rm TIR}$) between $L_{\rm IR, SF}$ and $P_{\rm 1.4\, GHz}$.
    \label{tab:properties1}}
    \tablehead{Property & & Value}
    \startdata
    RA & (J2000) & 34.35823432 \\
    Dec & (J2000) & -5.24091976 \\
    $z_{\rm phot}$ & & $3.78_{-0.03}^{+0.01}$ \\
    $f_{\rm AGN}$ & $(3-30\,\rm \mu m)$ & $0.90\pm 0.01$ \\
    $F_{\rm 870\,\mu m}$ & $(\rm mJy)$ & $3.59 \pm 0.64$ \\
    $\log{L_{\rm 6\, \mu m, AGN}}$ & $(\ergs)$ & $46.15 \pm 0.02$ \\
    $\log{L_{\rm 6\, \mu m, tot}}$ & $(\ergs)$ & $46.19 \pm 0.02$ \\
    $\log{L_{\rm 2-10\, keV}}$ & $(\ergs)$ & $< 44.03$ \\
    $\log{L_{\rm acc}}$ & $(\ergs)$ & $47.12 \pm 0.58$ \\
    ${\rm BHAR(\epsilon = 0.1)}$ & $(\Msunyr)$ & $20.4 \pm 3.1$ \\
    $\log{M_\star}$ & $(\Msun)$ & $10.82 \pm 0.57$ \\
    ${\rm SFR_{SED}}$ & $(\Msunyr)$ & $157 \pm 8$ \\
    $\log{L_{\rm IR}}$ & $(\Lsun; \rm 8-1000\, \mu m)$ & $13.25 \pm 0.16$ \\
    $\log{L_{\rm IR,SF}}$& $(\Lsun; \rm 8-1000\, \mu m)$ & $12.11 \pm 0.16$ \\
    $\log P_{1.4\,\rm GHz}$ & $(\rm W\, Hz^{-1})$ & $25.46 \pm 0.03$ \\
    $q_{\rm TIR}$ & & $0.68 \pm 0.16$
    \enddata
\end{deluxetable}

\begin{figure*}
    \centering
    \includegraphics[width=\linewidth]{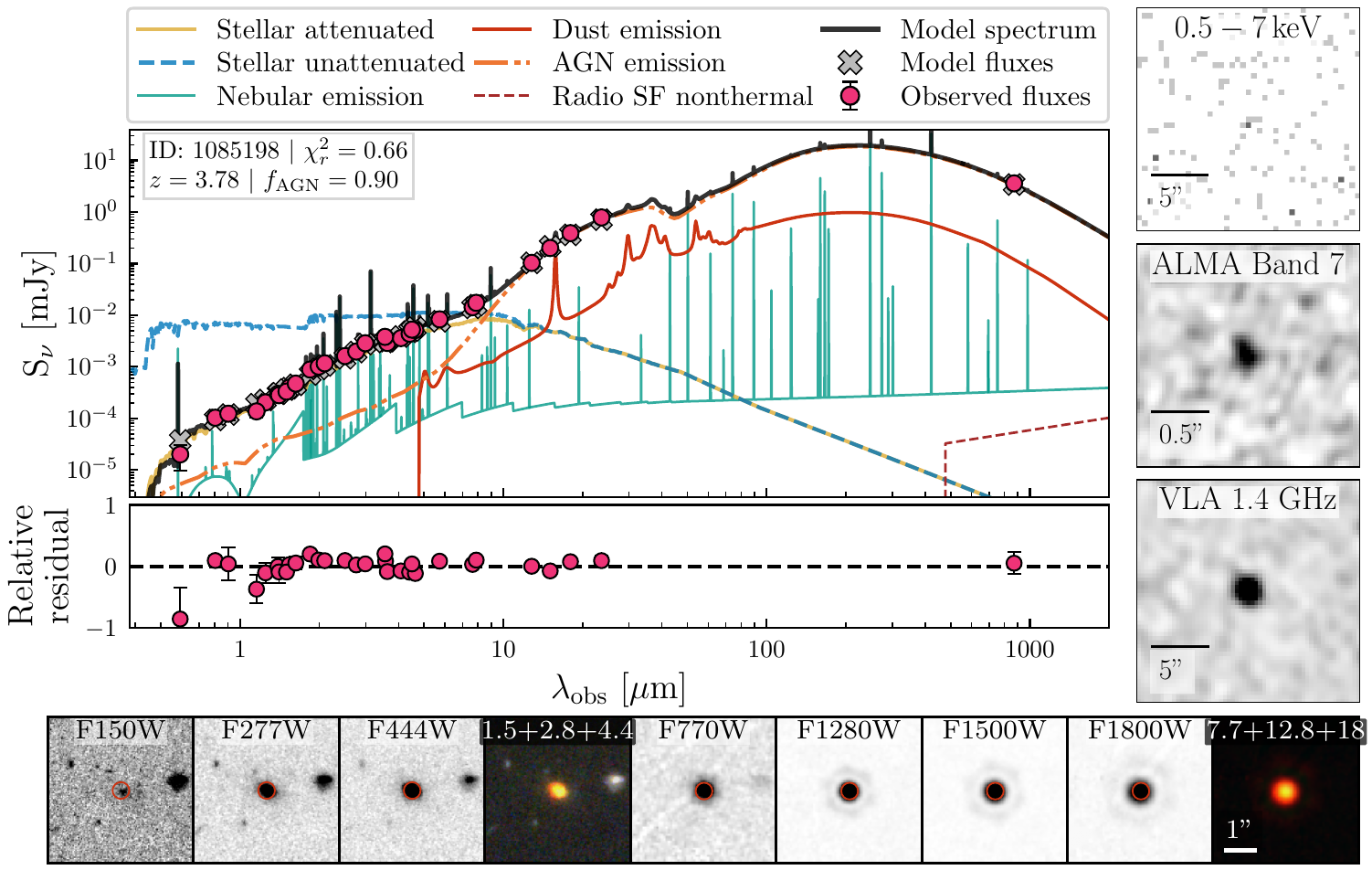}
    \caption{The best-fit SED from \cigale and image cutouts in \textit{JWST} filters, \textit{Chandra} $0.5-2\,\rm keV$, ALMA Band 7, and VLA 1.4 GHz. We also present a NIRCam RGB image using F150W (1.5 $\mu$m), F277W (2.8 $\mu$m), and F444W (4.4 $\mu$m); and a MIRI RGB using F770W (7.7 $\mu$m), F1280W (12.8 $\mu$m), and F1800W (18 $\mu$m). Both the SED and the \textit{JWST} cutouts show a luminous mid-IR AGN and dust-attenuated starlight in the rest-UV and optical. The ALMA Band 7 map shows compact cold dust emission, and the AGN is undetected by \textit{Chandra}. The mid-IR brightness combined with the inferred X-ray flux upper limit suggest CT obscuration. We also find a radio excess at 1.4 GHz with our updated measurement of the total IR luminosity. This excess is likely AGN-driven, meaning that \one may be a rare radio-loud \ctagn.}
    \label{fig:sed1}
\end{figure*}

We present the best-fit \cigale SED from the rest-UV to the far-IR alongside image cutouts from X-ray to radio in \figref{sed1}. The SED fit to the rest-frame mid-IR (probed by MIRI and \textit{Spitzer} $\rm 24\, \mu m$) provides strong evidence of hot dust emission as found in obscured AGN \citep{Hickox18}, with an estimated AGN fraction of $\approx 90\%$ over $3-30\,\rm \mu m$. The inferred AGN fraction is similarly large over the full IR. As is typical for buried AGN in dusty galaxies we see faint, attenuated starlight in the rest-UV and optical, and the absorbed starlight is reprocessed as IR dust emission. Observations of the far-IR spectral peak are necessary to study the extent to which the observed cold dust emission is driven by the AGN as opposed to star formation. The observed X-ray flux is below the detection limit from the background, while the source is strongly detected at $1.4\,\rm GHz$. The half-light radius in F200W from the morphology catalog is $1.86^{+0.12}_{-0.12}$\,kpc, which is near the median for massive star-forming galaxies at $3 < z <4$ \citep{Chen26}. In F444W the half-light radius is $1.20^{+0.01}_{-0.02}$\,kpc, indicating that \one is not a little red dot as its rest-optical flux is more extended than the typical $<0.1\,\rm kpc$ for LRDs \citep{Baggen26, Cloonan26}.

The properties of \one and its host galaxy are measured directly from the \cigale SED fitting; the MIR AGN fraction ($f_{\rm AGN}$), the accretion power of the AGN ($L_{\rm acc}$), the stellar mass ($M_\star$) and star formation rate (SFR; over the last $100\Myr$) of the host galaxy, and the total IR luminosity of the source (galaxy$+$AGN) over $8-1000\,\rm \mu m$. The values of these properties are listed in \tabref{properties1}, along with basic information including sky coordinates and photometric redshift. The accretion power ($L_{\rm acc}$) provided by \cigale is defined as the intrinsic (i.e., corrected for absorption) luminosity of the accretion disk averaged over viewing angle.

\textit{El Ascua} is among the most extreme AGN at $z\sim 4$ which belong to the missing obscured population known from the cosmic X-ray background \citep[e.g.,][]{Ueda03} and from comparing MIR and X-ray luminosity functions \citep[LFs; e.g.,][]{Runburg22, Bulichi26}. Unsurprisingly, with $\log\left( L_{\rm acc}/\ergs \right) \sim 47$, it falls at the luminous end of the MIRI AGN LF at $z\sim 4$. However, the MIR and X-ray LFs at $z\sim 4$ begin to converge around $\log\left( L_{\rm acc}/\ergs \right) \sim 47$; in other words, the fraction of missing X-ray-faint AGN decreases as $L_{\rm acc}$ increases toward $\log\left( L_{\rm acc}/\ergs \right) \sim 47$ \citep{Bulichi26}. We argue that while \ctagn are likely common overall at $z\sim 4$, \ctagn of such high luminosity as \one are statistically rare.

Additionally, \one is of similar luminosity to other remarkably bright \ctagn at $z\sim 4$ \citep[e.g.,][]{Gilli14, Vito20, Villani26}, and it is slightly less luminous than hot dust-obscured galaxies \citep[hot DOGs;][]{Fan16, Wu18, Li24} and the rare `extremely red quasars' \citep{Ma24}. Compared to local \ctagn, \one is one or two orders of magnitude more luminous \citep[e.g.,][]{Spinoglio22, Spinoglio24, Annuar25}.

From $L_{\rm acc}$, we compute the black hole accretion rate (BHAR) following the formalism used by \citet{Marconi04} and \citet{Yang23}:
\begin{equation}\label{eqn:bhar}
    {\rm BHAR} = \frac{L_{\rm acc} (1 - \epsilon)}{\epsilon c^2}
\end{equation}
where $\epsilon$ is the radiative efficiency and $c$ is the speed of light. This formalism for BHAR is derived by subtracting the fraction of mass converted into radiation from the total accreted mass, assuming spherical accretion. For \one, we estimate $\rm BHAR = 20.4\pm 3.1\, \Msunyr$ assuming a population-average $\epsilon \approx 0.1$. If we instead assume a theoretical maximum of $\epsilon \approx 0.42$ \citep{Thorne74, Ananna20} for a rapidly rotating Kerr black hole, we find $\rm BHAR = 3.23\pm 0.24\, \Msunyr$.

The SMBH in \one is growing more quickly than anticipated from the galaxy stellar mass and SFR. Even with the upper limit of $\epsilon \approx 0.42$ from theory, we find a large ratio of $\rm BHAR / SFR = 0.020 \pm 0.002$. For reference, \citet{Stone22} find a characteristic $\rm BHAR / SFR$ of $\sim 0.01$ for a stack of LIRGs at $z\sim 0$ with $f_{\rm AGN} \approx 0.96$. A related metric, the specific BHAR (sBHAR), paints the same picture and can be defined as a modified ratio of AGN accretion power to galaxy stellar mass \citep{Habouzit19}:
\begin{equation}
    {\rm sBHAR} = \frac{L_{\rm acc}}{10^{38}\,\ergs} \times \left( \frac{0.002M_\star}{M_\odot} \right)^{-1}.
\end{equation}
We find $\rm sBHAR = 9.95\pm 1.1$ for \one. This value falls on the tail of the sBHAR distribution function for star-forming galaxies of $10 < \log M_\star < 11$ at $3 < z < 4$, which has an average in the range of $\rm \braket{sBHAR} \sim 0.1-0.3$ and begins to decrease dramatically at $\rm sBHAR \gtrsim 0.5$ \citep{Aird18, Habouzit19}. Large $\rm BHAR/SFR$ and sBHAR values are often associated with accretion rates nearing or exceeding the Eddington limit \citep{Aird18}.

We make a rough estimate for the Eddington ratio $\lambda_{\rm Edd} \equiv L_{\rm acc}/L_{\rm Edd} \propto M_{\rm BH}^{-1}$ \citep{Marconi04} by assuming that \one falls within $3\sigma$ of the $M_{\rm BH}-M_\star$ scaling relation at $3 < z < 4$ \citep{Kormendy13, Greene20}. If we suppose an average mass ratio of $\braket{M_{\rm BH}/M_\star} \sim 10^{-3}$ with an intrinsic scatter of $\sigma_{\rm int} \approx 0.9\dex$, as measured by \citet{Jones25}, we find an extremely conservative $3\sigma$ upper limit of $\log\left( M_{\rm BH}/\Msun \right) < 10.52$ given a stellar mass of $\log\left( M_\star/\Msun \right) \approx 10.82$. Paired with the AGN accretion power, this upper limit corresponds to a conservative lower limit on the Eddington ratio of $\lambda_{\rm Edd} > 0.031$. We note that we cannot claim that the black hole is as remarkably massive as $\log\left( M_{\rm BH}/\Msun \right) \sim 10.5$, as more precise estimates of $M_{\rm BH}$ are not available with current data. Indeed, the mass is most likely significantly lower given the SMBH mass function. The MIRI imaging in UDS from PRIMER and MINERVA probes a comoving volume of $\approx 3.4\times 10^{6}\,\rm Mpc^{3}$ over $3.5 < z < 4.25$, whereas the mass function predicts $\sim 1$\,BH with $\log\left( M_{\rm BH}/\Msun \right) \sim 10.5$ over a much larger comoving volume of $\sim 10^9\,\rm Mpc^{3}$ in the same redshift bin \citep[e.g.,][]{He24}. We therefore anticipate that the Eddington ratio of \one is significantly higher than the lower limit we find. Follow-up spectroscopy with NIRSpec can measure $M_{\rm BH}$ and test that expectation.

From these various diagnostics of BH accretion, we conclude that \one might be nearing the peak of a phase of obscured BH growth before AGN feedback expels the obscuring material. Pairing the conservative lower limits on $\lambda_{\rm Edd}$ with the inferred $N_H > 10^{24}\,\cm^{-2}$, we find that \one lies near the end of `Stage 2' in the radiation-regulated unification model outlined by \citet{Ricci17a, Ricci23}. In this framework, SMBHs begin in a quiescent, low-$\lambda_{\rm Edd}$ state (Stage 1), and then accretion begins with an obscured phase where both $\lambda_{\rm Edd}$ and $N_H$ increase as the SMBH grows. Then, after surpassing the `effective' Eddington limit $\lambda_{\rm eff}(N_H)$ for the dusty gas (which is a function of column density), AGN radiative feedback becomes much more efficient. The AGN at this point enters a brief `blowout' phase (Stage 3) before transitioning into an unobscured quasar with high $\lambda_{\rm Edd}$ and low $N_H$ (Stage 4). Most hot DOGs and other hyper-luminous reddened quasars are in or nearing the blowout phase \citep[e.g.,][]{Wu18, Jun21, Kayal24, Zou25}. With \one located in the extreme end of Stage 2 but slightly less luminous than most hot DOGs, the AGN could be nearing the effective Eddington limit and therefore located in a pre-hot DOG phase.

\subsection{The radio emission and possible radio-loudness}

\begin{figure}
    \centering
    \includegraphics[width=\linewidth]{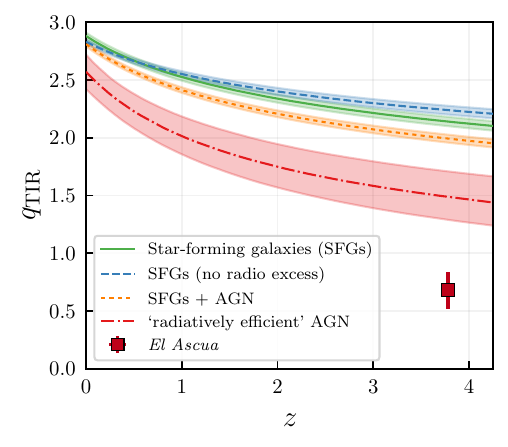}
    \caption{The observed ratio for \one between SF-driven IR luminosity and the 1.4\,GHz radio power, measured with \cigale, is consistent with an AGN-driven radio excess. We present empirical curves from \citet{Delhaize17} for the ratio $q_{\rm TIR}(z)$ as a function of redshift in different source populations. Star-forming galaxies are shown in blue and green, while all sources are shown in orange and `radiatively efficient' AGN are shown in red. We note that this population of radiatively efficient AGN are mostly radio-quiet \citep{Delvecchio17, Smolcic17}.}
    \label{fig:radio}
\end{figure}

Star-forming galaxies exhibit a strong correlation between IR and radio luminosities, and the ratio between them can provide an additional signature of AGN activity in the form of a radio excess \citep[e.g.,][]{DelMoro13, Delhaize17, Zhang25}. \citet{Sajina07} find that a large fraction ($\sim 40\%$) of dusty mid-IR AGN with silicate absorption at $9.7\rm\,\mu m$ are radio loud. We find that \one is strongly detected at 1.4 GHz with the VLA \citep{Simpson06}. To investigate whether \one is radio-loud or quiet, we measure the ratio $q_{\rm TIR}$ between the total IR and radio luminosities as:
\begin{equation}
    q_{\rm TIR} = \log \left( \frac{L_{\rm IR, SF}}{3.75 \times 10^{12}\,\rm Hz} \right) - \log \left( \frac{P_{1.4\,\rm GHz}}{\rm W\,Hz^{-1}} \right),
\end{equation}
where $L_{\rm IR, SF}$ is the IR luminosity in $\rm W$ due to star formation (from the SED fit), and $P_{1.4\,\rm GHz}$ is the radio power at 1.4 GHz in $\rm W\, Hz^{-1}$. We then compare to the redshift evolution $q_{\rm TIR}(z)$ for different source populations as measured by \citet{Delhaize17}, namely star-forming galaxies and `radiatively efficient' AGN. This population of AGN is a union of mid-IR-, X-ray-, and SED-selected AGN and they are mostly radio-quiet \citep{Delvecchio17, Smolcic17}. The results are shown in \figref{radio}.

We find that \one has a radio excess; i.e., its $q_{\rm TIR}$ value is significantly lower than predicted for luminous AGN at its redshift. The synchrotron radiation is likely driven by the AGN, as star formation does not account for the observed power at 1.4\,GHz. We conclude that \one is likely a radio-loud AGN, which matches the result from \citet{Zhang25} who include \one in their catalog of radio AGN identified via a similar technique with the IR\textendash radio correlation. Radio-loud \ctagn are notably rare with only a handful of confirmed sources \citep{LaMassa23}. If \one is indeed radio-loud, then it may be the most distant example yet known, and it would provide a novel opportunity to study the obscuring material in AGN in the distant universe.

\section{Summary}

In this work we present the discovery of a remarkably luminous \ctagn at $z\approx 3.78$ with multi-band MIRI imaging from the MINERVA survey \citep{Muzzin26} and archival X-ray data from the XUDS survey \citep{Kocevski18}. Our results and interpretations are summarized as follows:
\begin{itemize}
    \item We find that \one is among the most luminous mid-IR AGN in UDS, with $\log\left( L_{\rm 6\,\rm \mu m}/\ergs \right) \approx 46.15\pm 0.02$. Our X-ray measurement of $\log\left( L_{\rm 2-10\,\rm keV}/\ergs \right) < 44.03$ places \one on the $L_{\rm 6\,\rm \mu m}-L_{\rm 2-10\,\rm keV}$ scaling relation expected for AGN with gaseous obscuring material of column density $N_H \approx 10^{24}\,\rm cm^{-2}$ \citep{Fiore09, Gandhi09, Lansbury15}. This result suggests that the AGN is Compton-thick.
    \item From SED fitting with \cigale, we find that \one has a powerful accretion luminosity of $\log\left( L_{\rm acc}/\ergs \right) \approx 47.12 \pm 0.58$, at the extremely high end of the AGN luminosity function at $z\sim 4$ \citep{Bulichi26}. Given the host galaxy stellar masses and SFR, this luminosity corresponds to large $\rm BHAR/SFR$ and $\rm sBHAR$ values (and likely a high $\lambda_{\rm Edd}$ value as well) when compared to the AGN population at this redshift. These physical properties characterize \one as nearing or at a peak phase of heavily obscured SMBH growth, where accretion accelerates and the column density increases until powerful AGN feedback clears the obscuring material \citep{Ricci17a, Ricci23}. This phase is comparable to the stage of SMBH growth observed in hot DOGs \citep[e.g.,][]{Wu18, Zou25}.
    \item We measure the ratio between the IR and radio luminosities and find \one exhibits a radio-excess consistent with AGN-driven synchrotron radiation. An excess indicates that \one may be among the most distant radio-loud \ctagn known in the universe \citep{LaMassa23}.
\end{itemize}

Though located in the well-studied UDS field \citep{Lawrence07}, \one had not been flagged as a mid-IR-luminous AGN of any type previously. Its discovery with \textit{JWST} demonstrates the power that MIRI provides to the community for directly confirming powerful, heavily obscured AGN which we have long understood as missing from the census of SMBH growth.

Remarkably luminous sources such as \one are key laboratories for detailed study of heavily obscured AGN in the early universe. Follow-up NIRSpec and MIRI spectroscopy can corroborate the mid-IR evidence of AGN activity with emission line diagnostics of the narrow-line region \citep{Calabro23}, constrain the black hole mass with broad emission lines, and measure outflow kinematics to probe AGN feedback. In the further future, the far-IR PRIMA mission \citep{Glenn25} will measure the full IR SED with narrow-band photometry and provide rest-frame mid-IR spectroscopy at $z\sim 4$, directly measuring the accretion power \citep{Sajina22, Spinoglio22} and disentangling dust temperature components for heavily obscured AGN in the early universe. These rich, unprecedented IR datasets will map the connection between obscured black hole accretion and feedback \citep{Ricci17a, Ricci23} at high redshift. 

\begin{acknowledgments}

The authors note that no generative AI was used in the preparation of this manuscript, including the analysis described therein.

APC warmly thanks Pallavi Patil, Teodora-Elena Bulichi, Andrew Mizener, Leonardo Barba, and Catalina Zamora for helpful discussions, and they recognize the generous support of the National Science Foundation through the NSF Graduate Research Fellowship Program. APC also warmly acknowledges Aurelian Capello, who helped choose the name of the discovered object.

DM acknowledges generous support from the Leonard and Jane Holmes Bernstein Professorship in Evolutionary Science.

PD warmly acknowledges support from an NSERC discovery grant (RGPIN-2025-06182).

This work is based in part on observations made with the NASA/ESA/CSA James Webb Space Telescope obtained from the Space Telescope Science Institute, which is operated by the Association of Universities for Research in Astronomy, Inc., under NASA contract NAS 5-26555. Financial support for program JWST-GO-7814 is gratefully acknowledged and is provided by NASA through grants from the Space Telescope Science Institute, which is operated by the Association of Universities for Research in Astronomy, Incorporated, under NASA contract NAS 5-03127. The data described here may be obtained from the MAST archive at \dataset[doi:10.17909/cg7r-tq15]{https://doi.org/10.17909/cg7r-tq15}. Computational resources for this work were provided by the University of Massachusetts Amherst’s partnership with the Unity Research Computing Platform, a multi-institutional cluster led by the University of Massachusetts and the University of Rhode Island. The authors acknowledge the Tufts University High Performance Computing Cluster\footnote{\url{https://it.tufts.edu/high-performance-computing}}, which was utilized for the research reported in this paper. This research used the Canadian Advanced Network For Astronomy Research (CANFAR) operated in partnership by the Canadian Astronomy Data Centre and The Digital Research Alliance of Canada with support from the National Research Council of Canada, the Canadian Space Agency, CANARIE, and the Canadian Foundation for Innovation. The Cosmic Dawn Center is funded by the Danish National Research Foundation under grant DNRF140.

This research has made use of the NASA/IPAC Extragalactic Database, which is funded by the National Aeronautics and Space Administration and operated by the California Institute of Technology.

\end{acknowledgments}

\vspace{5mm}
\facilities{JWST/NIRCam, JWST/MIRI}

\software{\texttt{astropy} \citep{astropy13, astropy18, astropy22}, \numpy \citep{harris2020array}, \matplotlib \citep{Hunter:2007}, \scipy \citep{2020SciPy-NMeth}, \cigale \citep{Boquien19, Yang20}
          }

\bibliography{refs}{}
\bibliographystyle{aasjournalv7}

\end{document}